\documentclass[preprint,aps,onecolumn,floats,floatfix,amsmath,nofootinbib]{revtex4-1}
\usepackage{graphicx}
\usepackage{array,dcolumn}
\usepackage{calc,tabularx, epsfig,mathrsfs}
\usepackage{hyperref}
\usepackage{tikz}
\usepackage{amsmath,verbatim,enumerate}
\usepackage{amssymb}
\usepackage{multirow}
\usepackage{caption}
\usepackage{subcaption}

\usepackage{bm}
\usetikzlibrary{patterns}
\def\p{\partial}

\def\d{\delta}

\def\e{\epsilon}
\def\k{\kappa}

\def\om{\omega}

\def\O{\mathcal{O}}

\def\rp2{\mathbb{R}\mathrm{P}^2}
\newcommand{\non}{\nonumber}

\def\half{{\frac12}}

\newcommand{\bea}{\begin{eqnarray}}
\newcommand{\eea}{\end{eqnarray}}
\def\half{{\frac12}}
\def\d{\delta}
\def\k{{\omega,m}}
\def\x{\mathbf{x}}

\def\p{i\omega t +im\phi}
\def\j{ \sum_m \int d\omega}

\def \O{\mathcal{O}}

\def\be{\begin{equation}}
\def\ee{\end{equation}}

\def\bal{\begin{align}}

\def\eal{\end{align}}
\begin{document}

\title{Bulk reconstruction in rotating BTZ black hole}

\author{Nirmalya Kajuri}
\affiliation{Indian Association of Cultivation of Science, Jadavpur, Kolkata 700032}

\begin{abstract}
The bulk reconstruction program aims to obtain representations of bulk fields as operators in the boundary CFT. In this paper, we extend the program by obtaining the boundary representation for a scalar field in a rotating BTZ black hole. We find that the representation of the field near the inner horizon shows novel features. We also obtain a representation for fields inside the horizon as operators in a single boundary CFT using mirror operator construction. 
\end{abstract}

\maketitle
\section{Introduction}

According to the AdS/CFT conjecture, quantum gravity in $d+1$-dimensional asymptotically AdS spacetime and a conformal field theory living on the  $d$-dimensional boundary of the spacetime are equivalent. Therefore, one should be able to translate all of bulk physics to the boundary CFT such that one can answer all questions about bulk physics by carrying out calculations entirely in the boundary CFT. 

To be able to do this, one would have to translate bulk fields to operators in the boundary CFT. AdS/CFT correspondence in its original form does not directly tell us how to do this. The AdS/CFT extrapolate dictionary\cite{Balasubramanian:1998sn,Banks:1998dd,Balasubramanian:1998de} only tells us how to map the boundary value of a bulk correlation function to a correlator in the boundary CFT. 

Concretely, let us consider an asymptotically AdS geometry $g$  dual to a CFT state $|\psi_g\rangle$. Then for a scalar field $\phi$ dual to a CFT primary $\O$, the extrapolate dictionary tells us that:
\begin{align}
\label{gxp}\lim_{r\to\infty}r^{n\Delta}\langle\phi(r_1,\x_1)\phi(r_2,\x_2).....\phi(r_n,\x_n)\rangle_g=  \langle \psi_g|\mathcal{O}(\x_1)\mathcal{O}(\x_2)....\mathcal{O}(\x_n)|\psi_g\rangle
\end{align}

Here $\x_i$ are the boundary coordinates and $r,\x_i$ are the bulk coordinates.

This does not tell us to answer questions about physics deep inside the bulk from the boundary CFT. That could be done if one were able to construct a CFT operator $\phi_{CFT}$ which exactly mimics the bulk field, which is to say that it satisfies a relation like:

\begin{align}
\label{bulkrec}  \langle \phi (r_1,\x_1) ..\phi(r_n,\x_n) \rangle_{g} =  \langle \psi_g|\phi_{CFT} (r_1,\x_1).. \phi_{CFT}(r_n,\x_n)|\psi_g \rangle
\end{align}

The CFT operator $\phi_{CFT}$ is referred to as the boundary representation of the bulk field $\phi$.

The program of constructing such CFT operators for different fields and in different asymptotically AdS backgrounds is called the bulk reconstruction program\cite{ Dobrev:1998md, Bena:1999jv,Hamilton:2005ju,Hamilton:2006az,Heemskerk:2012np,Kabat:2011rz,Heemskerk:2012mn, Kabat:2012hp,Sarkar:2014dma,Kabat:2017mun,Kabat:2018pbj,Foit:2019nsr}). For a recent review, see \cite{Kajuri:2020vxf}. 

We will review how one can obtain boundary representations in the next section, for now we note that boundary representations have the form:

\be 
\phi_{CFT}(r,\x,\x') = \int \, d^d \x' K(r,\x;\x') \O(\x')
\ee

Here $K(r,\x;\x')$ is known as the smearing function. This construction is referred to as the HKLL construction in literature \cite{Hamilton:2005ju,Hamilton:2006az}. The boundary representation is a non-local operator in the CFT. As we can see here, knowledge of the smearing function $K(r,\x;\x')$ will allow one to translate any statement about bulk fields to statements about boundary operators.

The bulk reconstruction program has had many important applications and has led to several new insights. Paradoxes that appeared from comparing boundary representations of fields in overlapping AdS/Rindler wedges led to the conjectured connection between holography and quantum error correction \cite{Almheiri:2014lwa} (although a different resolution has also appeared \cite{Kajuri:2021vkg}), which has proved to be a productive line of enquiry. Another striking insight came from the observation that it is impossible to obtain boundary representations of horizon-crossing Wilson lines that connect the two boundaries in a two-sided AdS black hole. This observation led to the proposal that gauge fields must be emergent and a surprising connection with the weak gravity conjecture was found \cite{Harlow:2015lma}. 

A very fruitful avenue for the application of bulk reconstruction has been in black hole physics \cite{Hamilton:2006fh,Lowe:2008ra,Leichenauer:2013kaa,Guica:2014dfa,Roy:2015pga,Kabat:2016rsx}, especially in the context of the firewall paradox \cite{Heemskerk:2012np,Almheiri:2013hfa}. The question of whether an infalling observer will encounter a firewall can be directly translated to a bulk reconstruction question: of whether one can find boundary representations of bulk fields inside the horizon. This has led to several interesting ideas. A particularly interesting proposal is the mirror operator construction of Papadodimas and Raju\cite{Papadodimas:2012aq,Papadodimas:2013jku,Papadodimas:2015jra} (see \cite{Raju:2020smc} for a recent review) which shows that one can indeed reconstruct fields behind the horizon as CFT operators, but these operators will depend on the choice of black hole microstate. We will return to mirror operators later in the paper.

Another question of black hole physics, which is particularly puzzling from the point of view of AdS/CFT, is the issue of understanding what happens near inner horizons of rotating or charged black holes. For such black holes, it is possible to analytically extend the metric beyond the inner horizon. The evolution of fields is not unique beyond the inner horizon (as it depends on boundary conditions at the timelike infinity), so the presence of inner horizons signals a breakdown of determinism. For Kerr and Reissner-Nordstrom black holes, this problem is solved by the fact that the inner horizon is unstable\cite{Simpson:1973ua,Poisson:1990eh,Dafermos:2003wr}. Thus the principle of Cosmic Censorship i.e the statement that singularities are always hidden by horizons holds for these black holes. However, for rotating BTZ black holes the question of stability is not yet settled \cite{Dias:2019ery,Papadodimas:2019msp,Emparan:2020rnp,Hartnoll:2020rwq,Bhattacharjee:2020gbo}.

What makes rotating BTZ black holes a puzzle from the perspective of AdS/CFT is the fact that the analytically extended spacetime continues infinitely in the past and future. Therefore one might think that the dual to this spacetime is also an infinite product of CFTs. However, this cannot be the case because the different boundaries are time-like separated and cannot be independent and moreover the correct dual is known to be a thermofield double state\cite{Maldacena:2001kr}. So it becomes important to understand the inner horizon (and the possibility of going beyond it) from a CFT perspective. Then one would be able to understand how much of the maximally extended spacetime is dual to the pair of CFTs. There have been several attempts to relate the behavior of fields near the inner horizon to the properties of CFT correlators\cite{Levi:2003cx,Balasubramanian:2004zu,Balasubramanian:2019qwk}. 

In this paper, we take the first step towards translating questions about the inner horizon directly to questions about operators in the boundary CFT by carrying out bulk reconstruction in a rotating BTZ background. With the boundary representation of bulk objects in hand, one can begin to understand the behavior of fields near the inner horizon directly from the boundary CFT.

As we will review in the next section, smearing functions for fields in black hole backgrounds are distributions rather than functions. Therefore, one has to smear over fields and construct boundary representations for wave packets. This is what we have done in this paper, following the HKLL construction. In the regime where the wave packet is of high frequency, we obtained plots for the smearing function near the outer and inner horizons. 

We found novel features for the smearing function near the inner horizon. The support of the smearing function for wave packets near the inner horizon is qualitatively different from that of wave packets near the outer horizon. One way to interpret our result is by identifying the past and future inner horizons. It then appears that while the boundary CFT can 'see' points near the outer horizon through the outer horizon only, it can see the points near the inner horizon both through the outer and inner horizons. We will make this statement precise later. We note here that this observation might hold some clues for the resolution of the questions posed earlier. 

We have also considered a mirror operator construction for representing bulk fields near the outer and inner horizons. As we will review in the next section, mirror operator representations allow one to represent bulk fields beyond the horizon as operators on a single CFT. The boundary representations thus obtained can be used to study the behavior of fields near the inner horizon from a single CFT. 

With both the usual HKLL representation and the mirror operator representation in hand, one now has all the tools available to map bulk fields to boundary operators. One can then start tackling questions about the inner horizon. For instance, one can ask how fields near the inner horizon behave if it is perturbed by a shock-wave. In the last section, we will outline some concrete directions one can pursue in this regard.

In the next section, we briefly recall the basics of spinning BTZ black holes and those of bulk reconstruction in black holes. The third section presents our results. We conclude with a summary.

\section{Preliminaries}

\subsection{Spinning BTZ black hole}

The rotating BTZ metric is given by: 
\begin{equation}
\label{metric}
ds^2 = - f(r) dt^2 +\frac{1}{f(r)}dr^2 + r^2 (d\phi -\Omega dt)^2 
\end{equation}

where 
$$f(r) = \frac{(r^2-r_+^2)(r^2-r_-^2)}{r^2}$$

The spacetime has two horizons: $ r=r_- $ is the inner Cauchy horizon  and $ r=r_+ $ is the outer event horizon. The causal structure is shown in the figure.

 \begin{figure}
\centering
\includegraphics[width=50 mm, height= 50 mm]{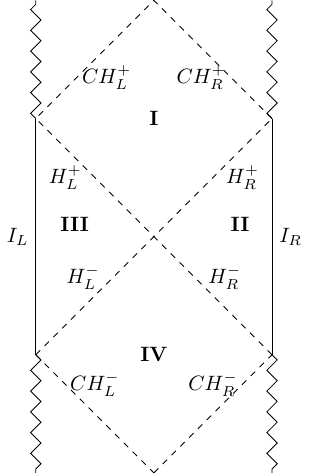}
\caption{Causal structure of a spinning BTZ black hole. $I_R$ and $I_L$ are the two boundaries. $H_L^+$ aand $H_L^-$ are the future and past parts of the left outer horizon, similarly $H_R^+$ aand $H_R^-$ are the future and past right outer horizon. $CH^+$ and $CH^-$ denote the future and past inner horizons. }
\end{figure}

The surface gravities of the two horizons are given by:

\be
\kappa_{\pm} =\frac{r_+^2 - r_-^2}{r_\pm}.
\ee

while the corresponding angular velocities are given by:

\be 
\Omega_\pm =\frac{r_\mp}{r_\pm}.
\ee

We now want to solve the Klein-Gordon equation for a massless field in this background:
\be 
\Box \phi = 0
\ee

The radial coordinate z will be useful to describe the solutions.

\be 
z= \frac{r^2-r_-^2}{r_+^2-r_-^2}
\ee

In these coordinates the inner horizon is at $z=0$, the outer horizon is at $z=1$ and the boundary is at $z= \infty$.

Using the symmetries of the metric we may write the solution as:

\be 
\label{ansat}
\phi = e^{i \omega t} e^{i l \phi} G(z)
\ee

Substituting \eqref{ansat} in the Klein-Gordon equation we obtain the hypergeometric equation in the radial coordinate:

\be 
z(1-z)F''(z) + [c-z(a+b+1)]F'(z) - abF(z)=0
\ee

where $F(z) = z^{-i \frac{\omega-m\Omega_-}{2\kappa_-}}(1-z)^{-i \frac{\omega-m\Omega_+}{2\kappa_+}}$ and
\begin{align}
a &= \frac{1}{2}\left( 2 - i \frac{\omega - m\Omega_-}{\kappa_-}-i\frac{\omega - m\Omega_+}{\kappa_+}\right)\\
b &= -\frac{1}{2}\left( i \frac{\omega - m\Omega_-}{\kappa_-}+i\frac{\omega - m\Omega_+}{\kappa_+}\right)\\
c&=1 - \frac{\omega - m\Omega_-}{\kappa_-}
\end{align}

The hypergeometric function has singular points at 0, 1 and $\infty$ and near each singular point, there is a convenient choice of basis. Of course, one can transform between the bases.
 
Near the boundary, the solutions are:
\begin{align}
\label{norm}
G_{\omega,m}^{norm,\infty} &= z^{-\frac{1}{2}(2a-c+1)}(z-1)^{\frac{1}{2}(a+b-c)} \hphantom{a}_2F_1\left(a,a-c+1;a-b+1;\frac{1}{z}\right), \\
G_{\omega,m}^{non-norm,\infty}&= z^{-\frac{1}{2}(2b-c+1)}(z-1)^{\frac{1}{2}(a+b-c)}\, _2F_1\left(a,b-c+1;-a+b+1;\frac{1}{z}\right)
\end{align}

Out of these, $G_{\omega,m}^{norm}$ is the normalizable mode. It falls off as $z^{-1}$ near the boundary. This is the one which we need for bulk reconstruction. 

Near the outer horizon, the convenient basis for the solutions is:

\begin{align}
\label{rp}
G_{\omega,m}^{R,+} &= z^{-\frac{1}{2}(1-c)}|1-z|^{-\frac{1}{2}(a+b-c)}\, _2F_1(c-b,c-a;-a-b+c+1;1-z) \\
\label{lp}
G_{\omega,m}^{L,+} &= z^{-\frac{1}{2}(1-c)}|1-z|^{\frac{1}{2}(a+b-c)}\,_2F_1(a,b:a+b-c+1;1-z)
\end{align}

At $r=r_+$ this basis behaves as:
\begin{align}
\label{rpa}
G_{\omega,m}^{R,+}|_{z \sim 1} &\sim |1-z|^{i\frac{\omega - l\Omega_+}{2\kappa_+}} \\
\label{lpa}
G_{\omega,m}^{L,+}|_{z \sim 1} &\sim |1-z|^{-i\frac{\omega - l\Omega_+}{2\kappa_+}}
\end{align}

The ``tortoise" radial coordinate $r_*$ is useful to describe near horizon solutions and we will use it (rather than $z$) when we work out smearing functions near the horizon in the next section:

\be 
r_*= \frac{1}{2 \kappa_+} \log \left| \frac{r-r_+}{r+r_+}\left(\frac{r+r_-}{r-r_-}\right)^{r_-/r_+}\right|
\ee

In terms of $r_*$, the near horizon behavior of these modes are:
\begin{align}
\label{rp22}
G_{\omega,m}^{R,+}| &\sim  \left( \frac{4r_+}{r_+^2-r_-^2}\left| \frac{r_+-r_-}{r_++r_-}\right|^{\frac{r_-}{r_+}}\right)^\frac{i(\om-m\Omega_+)}{\kappa_+} e^{i (\om-m\Omega_+) r_*}\\
\label{lp22}
G_{\omega,m}^{L,+}&\sim  \left( \frac{4r_+}{r_+^2-r_-^2}\left| \frac{r_+-r_-}{r_++r_-}\right|^{\frac{r_-}{r_+}}\right)^\frac{-i(\om-m\Omega_+)}{\kappa_+}e^{-i (\om-m\Omega_+) r_*}
\end{align}

These solutions behave like right and left moving waves near the horizon (hence the R/L labels). 

Near the inner horizon, the solutions are:

\begin{align}
\label{rm}
G_{\omega,m}^{R,-} &= z^{-\frac{1}{2}(1-c)}(1-z)^{\frac{1}{2}(a+b-c)}\, _2F_1(a,b;c;z) \\
\label{lm}
G_{\omega,m}^{L,-} &= z^{\frac{1}{2}(1-c)}(1-z)^{\frac{1}{2}(a+b-c)}\, _2F_1(a-c+1,b-c+1;2-c;z)
\end{align}

The near-horizon behavior of the two solutions is given by:

\begin{align}
\label{rma}
G_{\omega,m}^{R,-}|_{z \sim 0} &\sim |1-z|^{-i\frac{\omega - l\Omega_-}{2\kappa_-}} \\
\label{lma}
G_{\omega,m}^{L,-}|_{z \sim 0} &\sim |1-z|^{i\frac{\omega - l\Omega_-}{2\kappa_-}}
\end{align}
Again, in terms of the tortoise coordinate, they are given by:
\begin{align}
\label{rm2}
G_{\omega,m}^{R,-}| &\sim   \left( \frac{4r_-}{r_+^2 - r_-^2}\left| \frac{r_+-r_-}{r_++r_-}\right|^{\frac{r_+}{r_-}}\right)^\frac{i(\om-m\Omega_+)}{\kappa_-}e^{i (\om-m\Omega_-) r_*} \\
\label{lm2}
G_{\omega,m}^{L,-}| &\sim   \left( \frac{4r_-}{r_+^2 - r_-^2}\left| \frac{r_+-r_-}{r_++r_-}\right|^{\frac{r_+}{r_-}}\right)^\frac{-i(\om-m\Omega_+)}{\kappa_-}e^{-i (\om-m\Omega_-) r_*} 
\end{align}

One can go from one basis to the other. Later we will need to know how the normalizable mode at the boundary can be written in terms of the basis we used near the outer horizon. This is given by:
\be
G_{{\omega,m}^norm,\infty} = A(\omega,m)G_{\omega,m}^{R,+}+ B(\omega,m)G_{\omega,m}^{L,+}
\ee 

where 

\begin{align}
\label{basechange1}
A(\omega,m)=\frac{\Gamma(a-b+1)\Gamma(a+b-c)}{\Gamma(a)\Gamma(a-c+1)} \\
B(\omega,m) = \frac{\Gamma(-a-b+c)\Gamma(a-b+1}{\Gamma(1-b)\Gamma(c-b)}
\end{align}
We will also need to know how to write the basis near the outer horizon in terms of the one near the inner horizon:

\begin{align} 
\label{basechange2}
G_{\omega,m}^{R,+} = C(\omega,m)G_{\omega,m}^{R,-} + D(\omega,m)G_{\omega,m}^{L,-}\\
G_{\omega,m}^{L,+} = \tilde{C}(\omega,m)G_{\omega,m}^{L,-} + \tilde{D}(\omega,m)G_{\omega,m}^{R,-}
\end{align}

where

\begin{align}
\label{coeff}
C(\omega,m) = \frac{\Gamma(1-c)\Gamma(1-a-b+c)}{\Gamma(1-a)\Gamma(1-b)}\\
D(\omega,m) = \frac{\Gamma(c-1)\Gamma(1-a-b+c)}{\Gamma(c-a)\Gamma(c-b)}\\
\tilde{C}(\omega,m) = \frac{\Gamma(c-1)\Gamma(a+b-c+1)}{\Gamma(a)\Gamma(b)}\\
\tilde{D}(\omega,m) = \frac{\Gamma(1-c)\Gamma(a+b-c+1)}{\Gamma(a-c+1)\Gamma(b-c+1)}\\
\end{align}
\subsection{Bulk reconstruction in two-sided black holes}

The aim of this section is to outline the general strategy of obtaining a CFT representation of a bulk field in a black hole background. But before that, let us briefly recall how this is done in pure AdS background. For convenience, we assume a 2+1 dimensional AdS. We consider a scalar field $\phi(r,t,\theta$ dual to a boundary primary $\O(t,\theta)$.

One starts by defining the following boundary operators:

\be \label{creation}
 \mathcal{O}_{\k} =  \int  dt \,d\theta \, e^{-i\omega t +i m \theta } \,  \mathcal{O}(t,\theta)
 \ee 
 \be \mathcal{O}_{-\omega,-m} =  \int  dt \,d\theta \, e^{i\omega t - i m \theta } \,  \mathcal{O}(t,\theta).
\ee

It can be checked that these operators satisfy:

\be \label{dog}
\langle 0|[ \mathcal{O}_{\k} ,\mathcal{O}_{\k'}]|0\rangle = N_{\k}\delta(\omega+\omega')\delta(m+m')
\ee

where $|0\rangle$ is the CFT vaccuum and $N_{\k}$ is a constant. Thus we see that within correlators, these operators satisfy the same algebra as bulk creation and annihiliation operators, up to the multiplicative constant $N_{\k}$ . Therefore we may write:

\begin{align} \label{ads}
a_{\omega,m} &= \frac{1}{N_{\om,k}} \O_{\k}\\
a^\dagger_{\omega,m} &= \frac{1}{N_{\k}} \O_{-\omega,-m}
\end{align}

Where this relation is understood to hold when the LHS appears inside a CFT correlator. Thus we have obtained a boundary representation of the bulk annihilation and creation operators corresponding to the field $\phi$. Now we can simply substitute them in the mode expansion of the field to obtain:

\be \label{rec}
\phi_{CFT}=\sum_{\k} g_{\k} \O_{\k} +  {g^*}_{\k} \O_{-\omega,-m}
\ee

Here $g_{\k}$ are the normalizable modes of the bulk field in pure AdS.

This is a boundary operator whose correlators will satisfy \eqref{bulkrec} for pure AdS by construction. Using \eqref{creation}, we can rewrite  \eqref{rec} in the following form:

\be 
\phi_{CFT}(r,t,\theta)= \int \, dt' d\theta' K (r,t,\theta;t',\theta')\O(t',\theta')
\ee

Here  $K (r,t,\theta;t',\theta')$ is called the smearing function. This construction is referred to as the HKLL construction. In what follows we will drop the prefix 'CFT' from $\phi_{CFT}$ and refer to the boundary representation as just $\phi$.

We now turn to bulk reconstruction in a two-sided black hole. There are now two CFTs living on either boundary. The boundary dual to a two-sided black hole is a thermo-field double state which is an entangled state of the two boundaries.

The steps of bulk reconstruction are similar. Once again we consider a scalar $\phi$ dual to a primary $\O$. The first step is checking that the same operators \eqref{creation} satisfy the same algebra for the thermal state \cite{Papadodimas:2012aq}. Then \eqref{ads} is still true for a field in this geometry. From here on we refer to $\mathcal{O}_{-\omega,-m}$ as $\dagger{\mathcal{O}}_{\k}$.

As before, one we have the annihilation and creation operators in hand, 
we only need the normalizable mode solutions of the Klein-Gordon equation to complete the construction.

First, let us consider the exterior of the black hole, ie regions I and III. So for instance, in region I one solves the Klein Gordon equation and chooses the modes which are normalizable:
\be
\lim_{r \to \infty} f^{norm}_{\omega,m} \propto r^{-\Delta} 
\ee 

Near the horizon the normalizable modes will behave as linear combinations of left-moving and right-moving modes $ f^{L}_{\omega,m} \approx e^{\p + \omega r_*} $ and  $ f^{R}_{\omega,m} \approx e^{\p - \omega r_*} $
\begin{equation}
f^{norm}_{\omega,m}\; \propto  \; f^{(L)}_{\omega,m} +   e^{- 2 i \d_{\omega,m}}  \, f^{(R)}_{\omega,m} 
\end{equation}

The CFT representation in region I is then obtained as: 
\begin{equation} 
\label{region1}
\phi^{I}(\x, r) =   \j \, \left[ \mathcal O_{\k}\, f^{norm}_{\k} (r,t,\theta) + \mathcal O_{\k}^\dagger\, (f^{norm}_{\k} (r,t,\theta))^*  \right]
\end{equation}

One can formally write down the smearing function as: 
\begin{equation}\label{smear1}
K(r,t,\theta;t',\theta') =  \j \,e^{i(\omega(t- t')- m(\theta-\theta'))} f^{norm}_\k(r,t,\theta)
\end{equation} 

A similar procedure can be followed for region III.

Now we consider the interior or region II. Again, we solve the Klein Gordon equation and obtain the mode solutions. However, now there are no boundary conditions, there are only matching conditions at the horizon. One obtains the mode solutions $ \chi^{L}_\k (r,t,\theta)$ and $ \chi^{R}_\k (r,t,\theta)$ which behave near the horizon as $  e^{\p - \omega r_*} $ and $  e^{\p + \omega r_*} $ respectively. Now we can use matching at the horizon. The solution inside the horizon must match with $f^{norm}_\k$ at the horizon between regions I and III, for instance. This gives $ \chi^{L}_\k (r,t,\theta) \sim e^{\p } f^{L}_{\omega,m}  $ (up to a scaling factor). Similarly $\chi^{R}_\k (r,t,\theta)$ is obtained by matching at the horizon between regions II and III. 

The absence of a boundary condition means that the number of independent modes inside the horizon is double that of outside the horizon. This is expected as independent modes move into region II from both regions I and III. 

The CFT representation is given by:
\begin{align}
\label{region2} 
\phi^{II}( r,t,\theta) = &\j  \left[ \mathcal O_\k\, \chi^{(L)}_{\k}(r,t,\theta)  + \tilde{ \mathcal  O}^\dagger_{\k} \, \chi^{(R)}_{\k}(r,t,\theta) +\mbox{h.c.} \right] .
\end{align}

Here, as later, we refer to the CFT operators on the right as $\tilde{ \mathcal  O}$.

This can be re-written as:
\begin{align} \label{smear2}
\phi^{II}( r,t,\theta)=\int \, dt' \, d\phi'    K_L(r,t,\theta;t',\theta')\, \mathcal{O}(t',\theta')  +\,\int \, dt'' \, d\phi''   K_R(r,t,\theta;t'',\theta'') \, \tilde{\mathcal{O}}(t'',\theta'')
\end{align}

where 
\be 
K_L(r,t,\theta;t',\theta')=  \j \,e^{-i(\omega t'- m\theta')} \chi^{(L)}_{\k}(r,t,\theta) + \mbox{h.c}
\ee
It has support on both the left and right boundaries. 

The expressions for smearing functions in \eqref{smear1}, \eqref{smear2} always diverge. This is a peculiarity of black hole backgrounds -- that the smearing function does not exist as a function\cite{Hamilton:2005ju,Hamilton:2006az,Leichenauer:2013kaa} but as a distribution\cite{Morrison:2014jha}. This is not a problem as such, as the correct correlators are still obtained from the boundary representation. To avoid divergences, one can construct a wave packet by smearing the field over a region in spacetime and obtain a convergent expression for the CFT representation of the wave packet\cite{Leichenauer:2013kaa,Guica:2014dfa}.\footnote{An alternative to using wave packets that works for non-rotating black holes is to Wick rotate to de-Sitter space. Here one obtains a representation in a complexified boundary\cite{Hamilton:2006az,Hamilton:2006fh}. This approach does not work in this case as Wick rotation yields a complex bulk metric.}.
 This is what we will do in the next section.

Now let us review the mirror operator construction. For the CFT state $|\psi\rangle$ dual to ta two-sided black hole, we can carry out the mirror operator construction starting from the following observation from \cite{Papadodimas:2019msp}:

\bea 
\tilde{O}_{\om,m}|\psi\rangle= e^{-\pi \omega} O^\dagger_{\om,m} |\psi\rangle \\
\tilde{O}^\dagger_{\om,m}|\psi\rangle= e^{\pi \omega} O_{\om,m} |\psi\rangle 
\eea

We note that this is a state-dependent equation. For this particular state, one may represent the creation and annihilation operators on the right CFT by the operators on the right-hand side of the equation. Therefore we can make these substitutions and represent the fields behind the horizon as operators on a single boundary CFT.

For instance, the expression of a field in region II as given by \eqref{region2} now becomes:

\be \label{state}
\phi^{II}(r,t,\theta) = \j  \left[ \mathcal O_\k\, \chi^{(L)}_{\k}(r,t,\theta) +\mbox{h.c.} + e^{\pi \omega}\mathcal{O}_{\k} \, \chi^{(R)}_{\k}(r,t,\theta) +e^{-\pi \omega}\mathcal{O}^\dagger_{\k} \, \left(\chi^{(R)}_{\k}\right)^*(r,t,\theta) \right] .
\ee

Which translates to: 

\begin{align} 
\phi^{II}(\x, r)=\int \, dt' \, d\phi'    \left(K_L(r,t,\theta;t',\theta') +  K_{mirror}(r,t,\theta;t',\theta')\right) \, \mathcal{O}_R(t'',\theta'')
\end{align}

where

\be 
K_{mirror} (r,t,\theta;t',\theta')=  \j \,e^{\pi \omega-i(\omega t'- m\theta')}  \chi^{(R)}_{\k}(r,t,\theta) +e^{-\pi \omega+i(\omega t'- m\theta')}\left( \chi^{(R)}\right)^*_{\k}(r,t,\theta)
\ee

Thus we see that we have obtained a boundary representation on a single boundary, albeit at the cost of state-dependence.

\section{Smearing function for spinning BTZ black hole}

In this section, we present our results. First, we present the usual HKLL representation where the fields in the interior are represented as a sum of the operators on the left and right boundaries. Then we present a `mirror operator'-like construction where the representation is on a single boundary. 

\subsection{HKLL representation of the smearing function}

Using the mode solutions in section II A and following the general strategy outlined in section II B, we can now obtain the boundary representation for bulk fields in a spinning BTZ. However, as we observed before, the smearing function obtained from \eqref{smear1} diverges. One needs to introduce wave packets to obtain a convergent answer.

Instead of considering the field at a point, we smear them using a wave packet:
 \begin{equation}
\Phi(r,t_0,\theta_0)= \int\, dt \, d\theta\, \xi^*_{\omega_0,t_0}\, \eta^*_{\theta_0,m_0} \phi^+(r,t,\theta) +h.c
\end{equation}

where $\phi^+$ denotes the positive frequency part of the field. We follow the wavepacket construction of \cite{Guica:2014dfa}:

\be
\xi_{\omega_0,t_0} = e^{-i\omega_0 (t-t_0)} \frac{\sin  \left(\epsilon\, (t-t_0)\right)}{\sqrt{\epsilon}(t-t_0)}\, ,\, \eta_{m_0,\theta_0} = e^{im_0 (\theta-\theta_0)} \frac{\sin  \left(\epsilon\, (\theta-\theta_0)\right)}{\sqrt{\epsilon}(\theta-\theta_0)}
\ee

This gives a wave packet centered around $t_0,\theta_0$.

Now using \eqref{region1} and \eqref{smear1} we can obtain the CFT representation of the wave-packet $\Phi$ in region I:
 \begin{equation}
\Phi^{I}(r,t_0,\theta_0) = \int dt' K_{\omega_0,m_0}(r,t_0,\theta_0;t',\theta')\, \mathcal{O} (t',\theta)
\end{equation}

where:
\bea
\label{bounsm}
K^I_{\om_0,m_0} (t_0-t,r,\phi_0-\phi)& = & \sum_m \frac{1}{(2\pi)^2}  \int d\om \, \left[  e^{i \om t- i m \phi} \tilde{\xi}^\star_{\om_0,t_0} (\om) \tilde \eta^\star_{m_0,\phi_0}(m) + \mbox{h.c.} \right] G_{\om,m}^{norm,\infty}(r) \non\\
& &\hspace{-2.5 cm} = \sum_{m=m_0 - \half \e}^{m_0+\half \e} \frac{2}{(2\pi)^2 \e}  \int_{\om_0 - \half \e}^{\om_0 + \half \e} d\om \, \cos \left[ \om (t_0-t) - m (\phi_0-\phi)\right] G_{\om,m}^{norm,\infty} (r).
\eea

Here $G^{norm,\infty}$ is the normalizable mode at the boundary given by \eqref{norm}.

Similarly from \eqref{region2} and \eqref{smear2} we obtain the CFT representation of the wave packet inside the horizon in region II:
 \begin{align}
\Phi^{II}(r,t_0,\theta_0) = \int dt' K^L_{\omega_0,m_0}(r,t_0,\theta_0;t',\theta')\, \mathcal{O}_L (t',\theta) + \int dt'' K^R_{\omega_0,m_0}(r,t_0,\theta_0;t'',\theta'')\, \mathcal{O}_R (t'',\theta'').
\end{align}

where: 
\bea
\label{outersm}
K^{L}_{0} (t_0-t,r,\phi_0-\phi)& = & \sum_m \frac{1}{(2\pi)^2}  \int d\om \, \left[  e^{i \om t- i m \phi} \, \tilde{\xi}^\star_{\om_0,t_0} (\om) \, \tilde \eta^\star_{m_0,\phi_0}(m) \, A(\om,m) G^{L,+}_{\om,m}(r) + \mbox{h.c.} \right]   \non\\
& &\hspace{-2.4 cm} = \sum_{m=m_0 - \half \e}^{m_0+\half \e} \frac{1}{(2\pi)^2 \e}  \int_{\om_0 - \half \e}^{\om_0 + \half \e} d\om \, \left[ e^{-i \om (t_0-t) + i m (\phi_0-\phi)}A(\om,m) G^{L,+}_{\om,m} (r) + \mbox{h.c.} \right].
\eea

where  $G^{L,+}_{\om,m}$ is given by the equation \eqref{lp}. A similar equation holds for the right smearing function.

To obtain a representation of the smearing function near the inner horizon we can use the basis of hypergeoemtric functions that is convenient in that region. To do this we use \eqref{basechange2} and obtain:

\begin{align}
\label{innersm}
\notag  K^{L}_{0} (t_0-t,r,\phi_0-\phi) =\sum_{m=m_0 - \half \e}^{m_0+\half \e} \frac{1}{(2\pi)^2 \e}  \int_{\om_0 - \half \e}^{\om_0 + \half \e} d\om \, \left[ e^{-i \om (t_0-t) + i m (\phi_0-\phi)}A(\om,m)\right. & \left. \left (\tilde{C}(\omega,m)G_{\omega,m}^{L,-}\right. \right.  \\+ \left. \tilde{D}(\omega,m)G_{\omega,m}^{R,-})\right)  &+\left.  \mbox{h.c.} \right]
\end{align}

where $G_{\omega,m}^{out,-}$ and $G_{\omega,m}^{in,-}$ are given by \eqref{rm} and \eqref{lm} respectively and $\tilde{C}(\omega,m)$ and $\tilde{D}(\omega,m)$ are given by \eqref{coeff}.

While these expressions of the smearing function are convergent, they cannot be written as closed-form expressions and offer little insight. One may obtain useful approximations near the outer and inner horizons.

First, we consider the field at a point in the exterior of the outer horizon which is close to the horizon. In the exterior region, the smearing function is given by \eqref{bounsm}. We can rewrite $G_{\om,m}^{norm,\infty}$ in terms of $G_{\om,m}^{R,+}$ and $G_{\om,m}^{L,+}$ using \eqref{basechange1}. Close to the outer horizon we can use \eqref{rp22} and\eqref{lp22}. Converting to the tortoise coordinate we finally get:

\be 
G_{\om,m}^{norm,\infty} \approx |a_{\om,m}| \cos(\om r_* +\delta_{\om,m})
\ee

where 

\begin{align}
|a_{\om,m}| =|A(\om,m)\alpha(\om,m)|^{1/2}\\
e^{2i \delta_{\om,m}}=\frac{A(\om,m) \alpha(\omega,m)}{B(\om,m) \alpha^*(\omega,m)}
\end{align}

where
\be 
\alpha(\om,m) = \left( \frac{4r_+}{r_+^2-r_-^2}\left| \frac{r_+-r_-}{r_++r_-}\right|^{\frac{r_-}{r_+}}\right)^\frac{i(\om-m\Omega_+)}{\kappa_+}
\ee

By choosing a wave packet in the high frequency regime $\om_0 \gg |m_0|\gg1$ we can obtain an expression for the smearing function. We use the following identities to simplify the formulae:
\bea 
\Gamma(ix)= \frac{\pi}{x \sinh x}\\
\Gamma(1+ix)=\frac{\pi x}{ \sinh x}
\eea

Further using the approximation which holds for $x \gg 1$:
\be 
i \log\left( \frac{\Gamma(ix)}{\Gamma(-ix)}\right) = 2x (\log x -1) -\frac{\pi}{2} +\mathcal{O}(x^{-1})
\ee

Using the above identities we get:
\bea 
|a_{\om,m}| \approx \frac{2}{\sqrt{2 \pi r_+}}\frac{r_+-r_-}{\om^{3/2}}\\
 \delta_{\om,m} \approx \frac{\pi}{4}
\eea

Then the smearing function for a high-frequency wave packet is given by:
\begin{align}
\notag K(r,t-t_0,\phi-\phi_0) \approx &\frac{2(r_+-r_-)}{\sqrt{2 \pi r_+}} \int\frac{d\om}{\om^{3/2}}   \left( cos [\om(t_0-t+r_*)-\pi/4\right.\\ &+\left.  m_0 (\phi_0- \phi)] +cos [\om(t_0-t-r_*)+\pi/4 + m_0 (\phi_0- \phi)]\right)
\end{align}

Here we have taken $\epsilon$ to be 1, thereby reducing the sum over $m$ to just a single $m_0$. Except for the prefactor, this agrees with the result derived for the non-rotating BTZ in \cite{Guica:2014dfa}. This can be integrated using Mathematica. The results are summarized in figure 2. 

\begin{figure}
\centering
\begin{subfigure}{.5\textwidth}
  \centering
  \includegraphics[width=.7\linewidth]{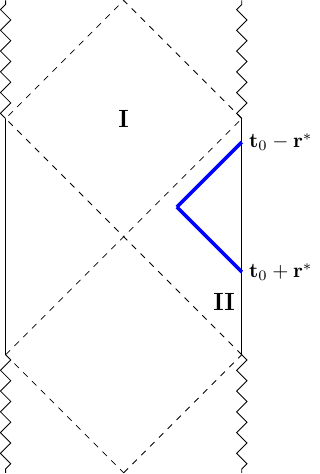}
  \caption{}
  \label{fig:sub1}
\end{subfigure}%
\begin{subfigure}{.5\textwidth}
  \centering
  \includegraphics[width=.9\linewidth]{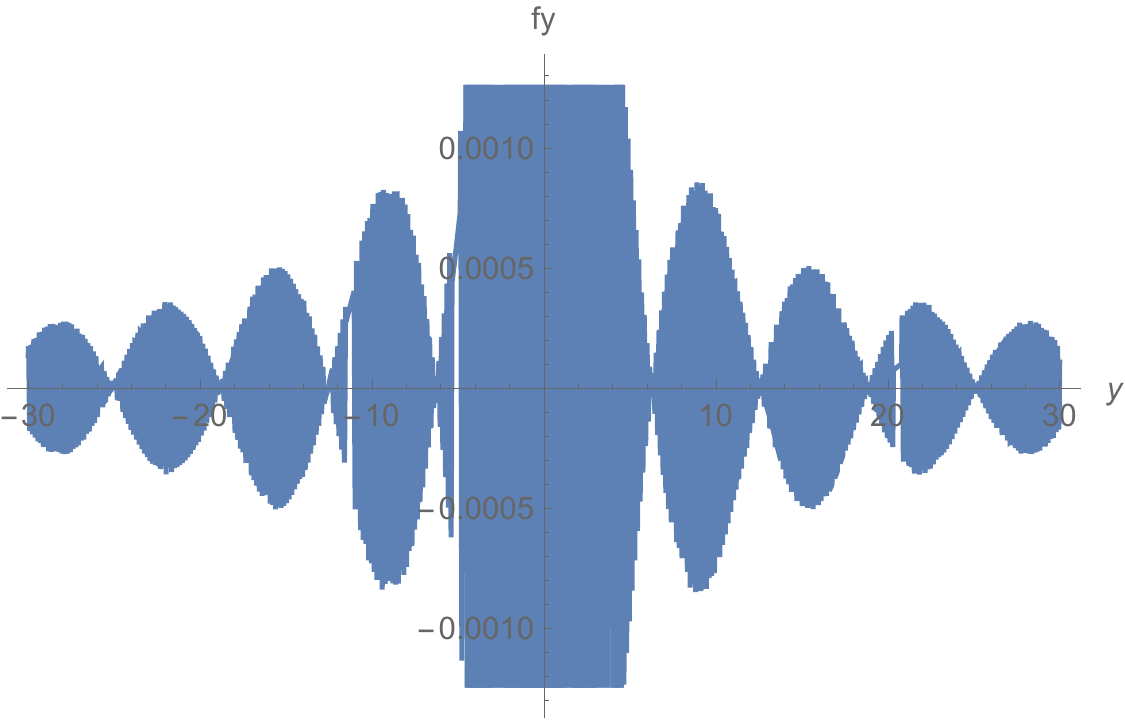}
  \caption{}
  \label{fig:sub2}
\end{subfigure}
\caption{(a) Bulk Reconstruction for wave packet in region I close to the outer horizon. The smearing function is peaked around two points $t_0+r_*$ and $t_0-r_*$ in one of the boundaries. (b) Plot of the smearing function for a point in the exterior near the outer horizon with boundary time. $\phi_0-\phi$ is taken to be zero. $\omega_0$ is chosen to be 40 and $\epsilon =1$.  }
\label{fig:test}
\end{figure}

The second case we may consider is that of a wave packet close to the outer horizon but in the interior of the black hole.

In this case, the smearing function is given by \eqref{outersm}. Once again we use \eqref{rp22} and the approximations above. The resulting expression for smearing function is given by:

\be 
K^L(r,t-t_0,\phi-\phi_0) \approx \frac{2(r_+-r_-)}{\sqrt{2 \pi r_+}} \int\frac{d\om}{\om^{3/2}}    cos [\om(t_0-t-r_*)+ m_0 (\phi_0- \phi)] 
\ee
 A similar expression holds for the $K^R$. 
 
\begin{figure}
\centering
\begin{subfigure}{.5\textwidth}
  \centering
  \includegraphics[width=.9\linewidth]{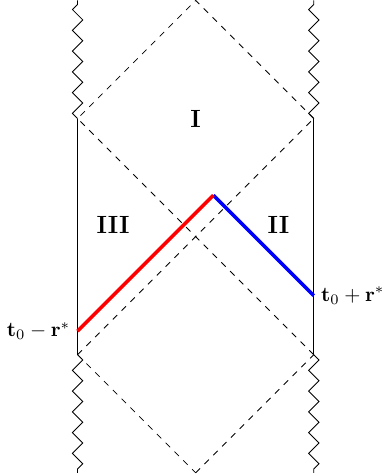}
  \caption{}
  \label{fig:sub1}
\end{subfigure}%
\begin{subfigure}{.5\textwidth}
  \centering
  \includegraphics[width=.9\linewidth]{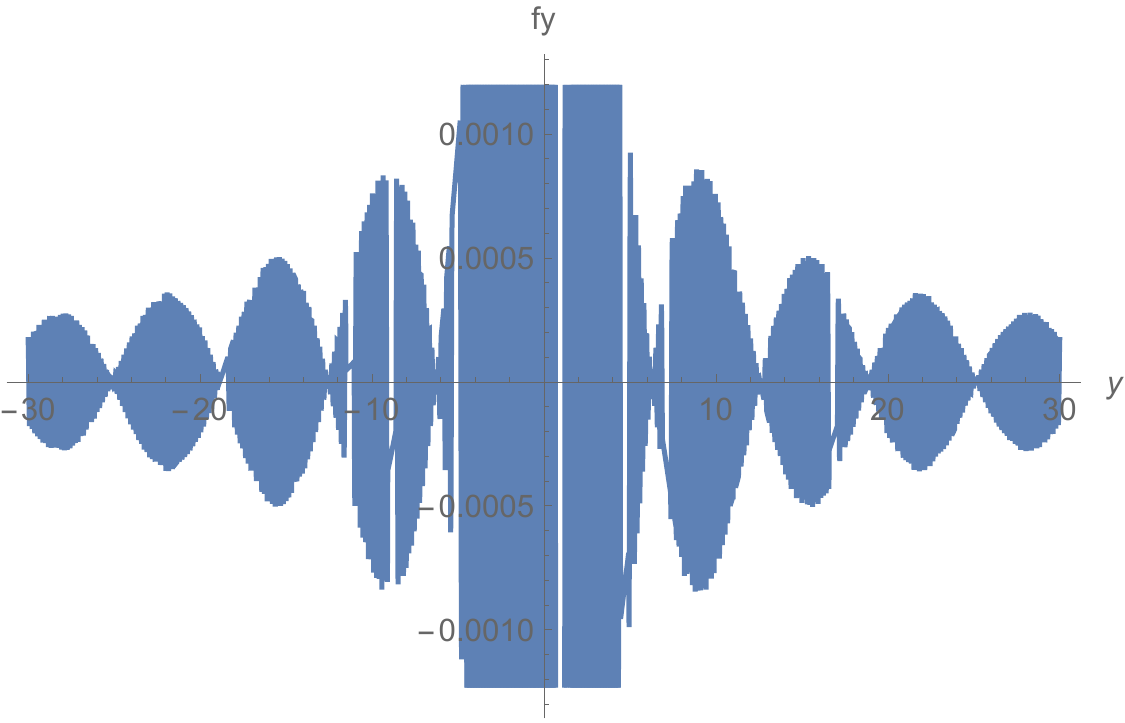}
  \caption{}
  \label{fig:sub2}
\end{subfigure}
\caption{(a) Bulk Reconstruction for wave packet in the interior close to the outer horizon. The smearing function is peaked around two points $t_0+r_*$ and $t_0-r_*$, one in each boundary. (b) Plot of the smearing function for a point in the interior near the outer horizon with boundary time. $\phi_0-\phi$ is taken to be zero. $\omega_0$ is chosen to be 40 and $\epsilon =1$. }
\label{fig:test}
\end{figure}

Finally, we consider a wave packet close to the inner horizon. In this case, the formula \eqref{innersm} applies for the smearing function. In this case, we have:

\be 
G^{out,-} = |C_n| \cos(\om r_* + \delta^{(1)}_{\om,m}) +|D_n| \cos(\om r_* + \delta^{(2)}_{\om,m})
\ee

where 
\bea 
|c_\k|= |A(\om,m)\tilde{C}(\om,m) \beta(\omega,m)|^{1/2}\\
|d_\k|= |A(\om,m)\tilde{D}(\om,m) \beta(\omega,m)|^{1/2}\\
 \delta^{(1)}_{\om,m}=\frac{\tilde{C}^*(\om,m) \beta^*(\omega,m) }{\tilde{C}(\om,m) \beta(\omega,m)}\\
 \delta^{(2)}_{\om,m}=\frac{\tilde{D}^*(\om,m) \beta^*(\omega,m) }{\tilde{D}(\om,m) \beta(\omega,m)}
\eea

where
\be 
\beta(\omega,m) = \left( \frac{4r_-}{r_+^2 - r_-^2}\left| \frac{r_+-r_-}{r_++r_-}\right|^{\frac{r_+}{r_-}}\right)^\frac{i(\om-m\Omega_+)}{\kappa_-}
\ee

Once again we use a high frequency wave packet. In this case we find that while the pre-factors are similar, the $\delta$ s turn out to be proportional to $\omega$. We write $\delta^{(1)}_{\om,m}= \om \Delta^{(1)} $ and $\delta^{(2)}_{\om,m}= \om \Delta^{(2)} $ where:

\bea 
\Delta^{(1)}=\frac{1}{2}\left( \frac{1}{r_++r_-}\log \frac{(r_+-r_-)r_+}{2r_-^2}+ \frac{1}{r_+-r_-} \log \frac{2r_+}{r_++r_-}\right)\\
\Delta^{(2)}=\frac{1}{2}\left( \frac{1}{r_++r_-}\log \frac{2r_+}{r_++r_-}+ \frac{1}{r_+-r_-} \log \frac{r_++r_-}{2r_+}\right)
\eea

Then the smearing function for a wave packet close to the inner horizon is given by:
\begin{align}
\notag K^L(r,t-t_0,\phi-\phi_0) \approx \frac{2(r_+-r_-)}{\sqrt{2 \pi r_+}} \int\frac{d\om}{\om^{3/2}} &  \left( \frac{\kappa_-}{\kappa_+}\cos [\om(t_0-t+r_*+\Delta^{(1)}) + m_0 (\phi_0- \phi)] \right. \\  & \left. +\frac{\kappa_+}{\kappa_-}\cos [\om(t_0-t-r_*-\Delta^{(2)}) + m_0 (\phi_0- \phi)]\right)
\end{align}

\begin{figure}
\centering
\begin{subfigure}{.5\textwidth}
  \centering
  \includegraphics[width=.9\linewidth]{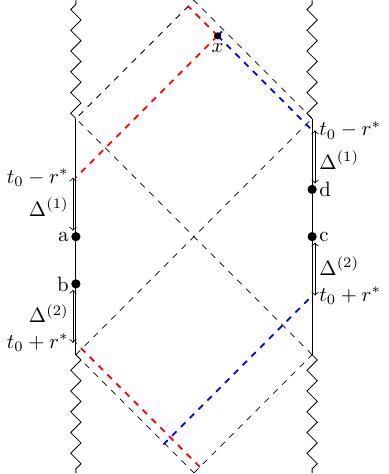}
  \caption{}
  \label{fig:sub1}
\end{subfigure}%
\begin{subfigure}{.5\textwidth}
  \centering
  \includegraphics[width=.9\linewidth]{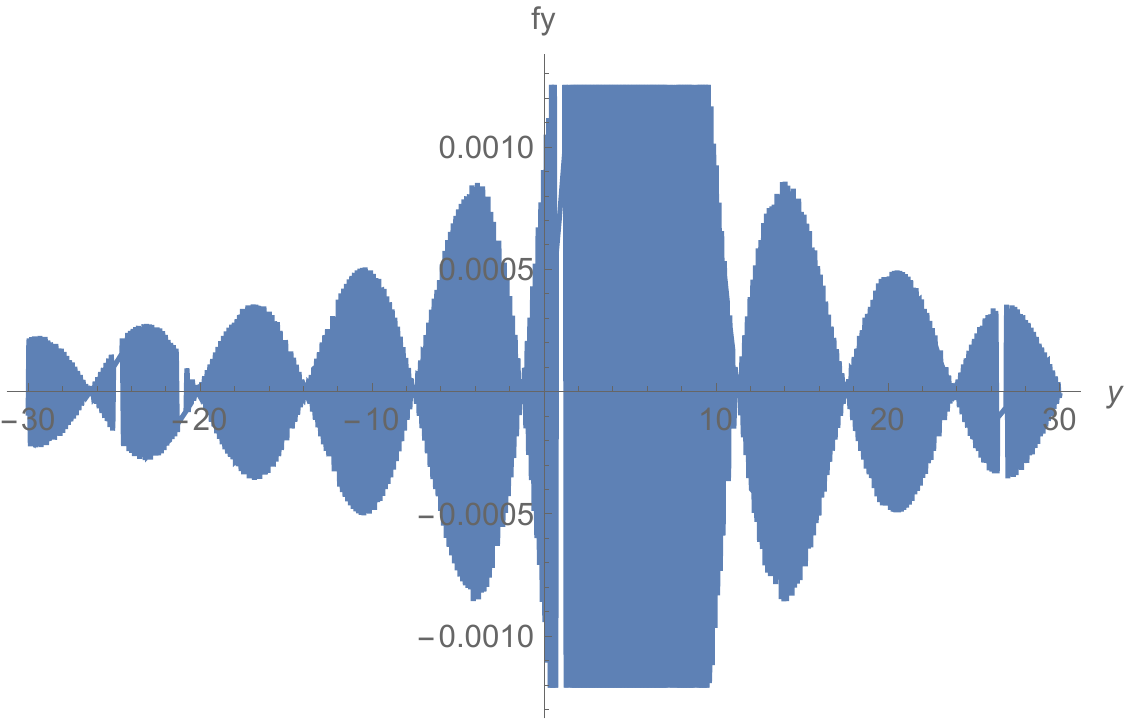}
  \caption{}
  \label{fig:sub2}
\end{subfigure}
\caption{(a) Bulk reconstruction construction for wave packet centered around the point $x$ in the interior close to the inner horizon. The smearing function is peaked around the four points a,b,c,d two at each boundary. (b) Plot of the smearing function for a point in the interior near the inner horizon with boundary time. $\phi_0-\phi$ is taken to be zero. $\omega_0$ is chosen to be 40 and $\epsilon =1$.}
\label{fig:test}
\end{figure}

We note two striking features in this last case. First, unlike previous cases, the smearing function is not exactly peaked at where the light ray reaches, but at a distance $\Delta^{(1)}$ or $\Delta^{(2)}$ from them. If the inner and outer horizons are close, this deviation can be significant. This is a rather surprising feature of the smearing function for points near the inner horizon, which demands further study.

Second, the smearing function is peaked on four points, two each on each boundary. Two of the points are similar to the ones for the outer horizon -- the peaks occur at points close to where the past light-ray from the bulk point reaches the boundary. But in this case, we get that the smearing function is peaked at two more points. These two points can be seen to be close to the ones where the future light ray from the bulk point would reach if the past inner horizon and the future inner horizon were to be identified. This novel feature also demands further study.

\subsection{Mirror operator representation of smearing function}

In the previous section, we obtained the HKLL representation of the smearing function which has support on both the boundaries. We can also use the mirror operator construction to obtain a representation which has support on only one boundary.

We can take \eqref{state} as our starting point. Once again, we consider a wave packet instead of a field point. The CFT representation of the wave packet inside the horizon then becomes:

\begin{align}
\Phi^{II}(r,t_0,\theta_0) = \int dt' \left(K^L_{\omega_0,m_0}(r,t_0,\theta_0;t',\theta')+K^{(mirror)}_{\omega_0,m_0}(r,t_0,\theta_0;t',\theta')\right) \mathcal{O}_L (t',\theta) 
\end{align}

where 

\begin{align}
\notag & K^{(mirror)} (t_0+t,r,\phi_0+\phi)
=\\ \notag & \sum_{m=m_0 - \half \e}^{m_0+\half \e} \frac{1}{(2\pi)^2 \e}  \int_{\om_0 - \half \e}^{\om_0 + \half \e} d\om  \left[ e^{-\pi \om} e^{-i \om (t_0+t) + i m (\phi_0+\phi)} G^{(L),+}_{\om,m} (r) + e^{\pi \om} e^{i \om (t_0+t) - i m (\phi_0+\phi)} \left(G^{(L),+}_{\om,m} (r)\right)^*  \right].
\end{align}

\begin{figure}
\centering
\begin{subfigure}{.5\textwidth}
  \centering
  \includegraphics[width=.7\linewidth]{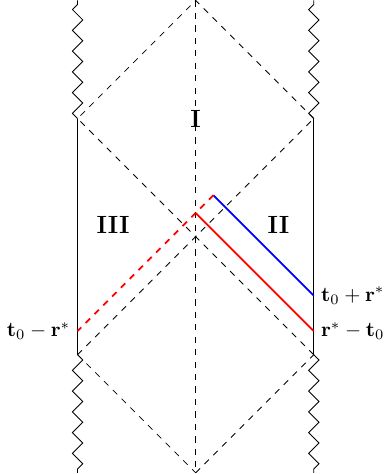}
  \caption{}
  \label{fig:sub1}
\end{subfigure}%
\begin{subfigure}{.5\textwidth}
  \centering
  \includegraphics[width=.9\linewidth]{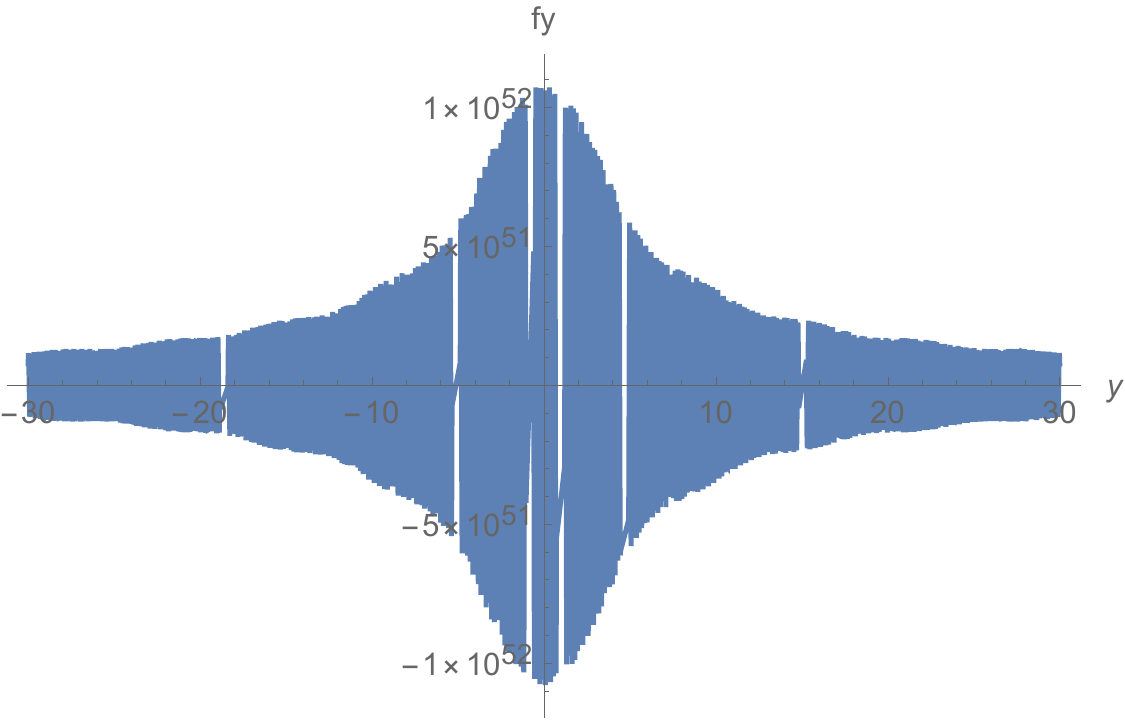}
  \caption{}
  \label{fig:sub2}
\end{subfigure}
\caption{(a) Mirror operator construction for wave packet in the interior close to the outer horizon. The smearing function is peaked around two points $t_0+r_*$ and $r_*-t_0$, both at the same boundary. The mirror point is obtained by reflecting back the red light ray at the center. (b) Plot of the real part of the mirror smearing function for a point in the interior near the outer horizon. $\omega_0$ is chosen to be 40 and $\epsilon =1$. The imaginary part is identical.}
\label{fig:test}
\end{figure}

Here we see that the peak of the mirror smearing function has a simple interpretation, it is obtained by reflecting the light ray connecting the bulk point to the left(or right) boundary (denoted by the dashed red line in the figure) from the center to the right boundary.

Using the approximations for a high-frequency wave packet close to the outer horizon we get the expression:
\begin{align}
\notag K^{(mirror)}(t_0+t,r,\phi_0+\phi) \approx \frac{2(r_+-r_-)}{\sqrt{2 \pi r_+}} \int\frac{d\om}{\om^{3/2}}   &( \cosh \pi \om \cos[ \om(t+t_0+r_*)+m_0(\phi+\phi_0) \\ &+ i \sinh \pi \om \sin [\om(t+t_0+r_*)+m_0(\phi+\phi_0)])
\end{align}
For a wave packet close to the inner horizon, using the same approximations one obtains:

 \begin{align}
\notag  K^{(mirror)}(t_0+t,r,&\phi_0+\phi)  \approx  \frac{2(r_+-r_-)}{\sqrt{2 \pi r_+}}  \int\frac{d\om}{\om^{3/2}}   \left( \frac{\kappa_-}{\kappa_+}e^{-\pi \om} +\frac{\kappa_+}{\kappa_-}e^{\pi \om}\right)\Big(\cos [\om(t_0+t+r_*+\Delta^{(1)})\\ \notag & +m_0 (\phi_0+ \phi)] +(\cos [\om(t_0+t-r_*-\Delta^{(1)}) +m_0 (\phi_0+ \phi)] \Big) + \left( \frac{\kappa_-}{\kappa_+}e^{-\pi \om} +\frac{\kappa_+}{\kappa_-}e^{\pi \om}\right)\\&\Big(\sin [\om(t_0+t+r_*+\Delta^{(2)}) +m_0 (\phi_0+ \phi)] +(\sin [\om(t_0+t-r_*-\Delta^{(2)}) +m_0 (\phi_0+ \phi)]\Big)
\end{align}

\begin{figure}
\centering
\begin{subfigure}{.5\textwidth}
  \centering
  \includegraphics[width=.7\linewidth]{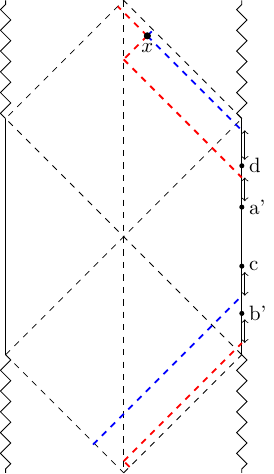}
  \caption{}
  \label{fig:sub1}
\end{subfigure}%
\begin{subfigure}{.5\textwidth}
  \centering
  \includegraphics[width=.9\linewidth]{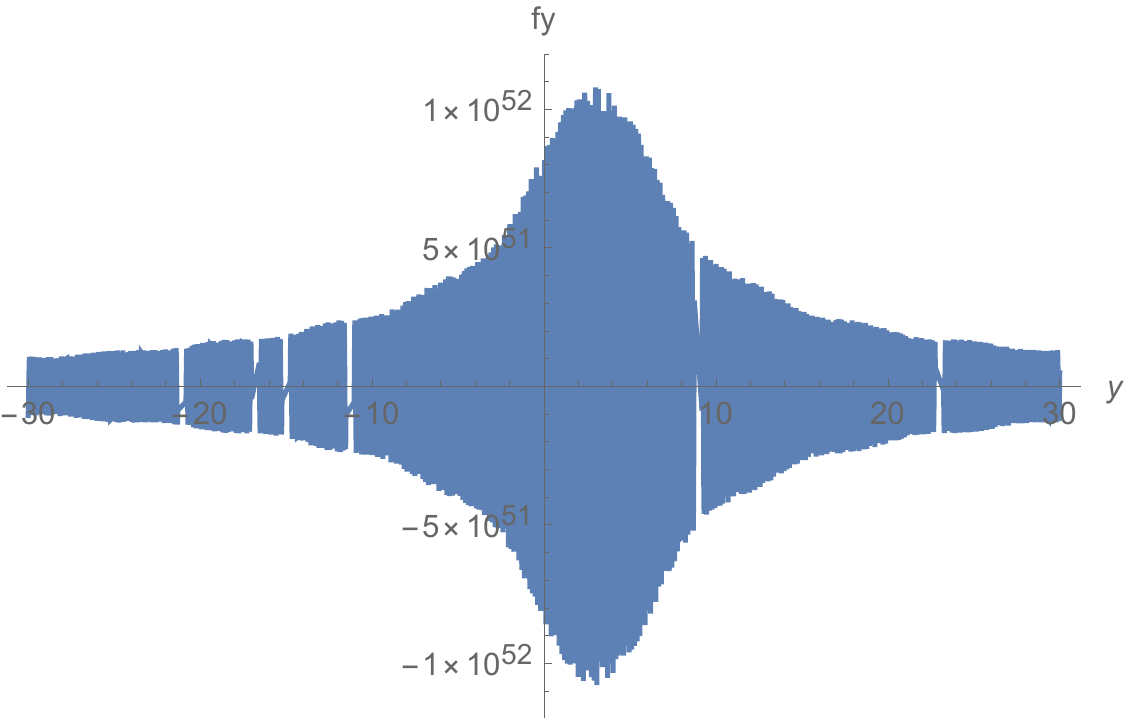}
  \caption{}
  \label{fig:sub2}
\end{subfigure}
\caption{(a) Mirror construction for wave packet centered around the point $x$ in the interior close to the inner horizon. The smearing function is peaked around the four points a',b',c,d on one boundary. a',b' are the mirrors of a and b in the previous diagram. (b) Plot of the real part of the mirror smearing function for a point in the interior near the inner horizon. $\omega_0$ is chosen to be 40 and $\epsilon =1$.}
\label{fig:test}
\end{figure}

Again the mirror operator smearing function is obtained by reflecting the left-moving rays from the center to the right boundary. Another interesting feature is that the mirror smearing function is complex. This bears further study in the future.

\section{Summary and Future directions}

In this paper, we carried out bulk reconstruction for a spinning BTZ black hole. We obtained boundary representations for a scalar field in both the exterior and interior of the horizon. Using high-frequency wave packets we obtained smearing functions near the inner and outer horizons. While the smearing function near the outer horizon had expected features, the wave packet near the inner horizon showed some novel features, which require further study. One feature is that the peaks are not exactly at the boundary points hit by light rays from the bulk, but are displaced from them in time. It is not clear what the implication of this is for translating the bulk theory to the boundary.
The other novel feature is that the smearing function is peaked around two points in each boundary, as opposed to one. The position of the second peak could be interpreted by identifying the past and future inner horizons and considering a light ray that passed through the future inner horizon and emerged from the past one. 

This feature might contain a clue about the questions about the inner horizon in AdS/CFT that we reviewed in the introduction. It appears that, in contrast to fields close to the outer horizon, the boundary 'sees' fields near the inner horizon not only through the outer horizon but 'through' the inner horizon as well. Put differently, a signal that seemed headed out to the timelike infinity seems to end up at the boundary CFT. These are, of course, speculative interpretations of the smearing function we have obtained, but they demand further investigation.

We also carried out a mirror operator construction for fields inside the horizon. This gives us a boundary representation of fields inside the horizon as operators on a single CFT. We saw that for high-frequency wave packets the peaks of the mirror operator smearing function on, say, the left boundary could be read off by reflecting the right moving rays from the center to the left boundary. We obtained mirror operators smearing functions for points close to the inner as well as outer horizons. One interesting feature was that the mirror operator smearing functions are complex. The implication of this is unclear and bears further study. 

In this paper we have translated, to the extent analytically possible, bulk fields in a rotating black hole background to operators in the boundary CFT. This can now be used to answer questions about the inner horizon directly from the boundary theories. We now outline some proposals about how one may proceed.

One possible avenue is to try to construct (possibly using a mirror operator construction) boundary representations for fields behind the inner horizon. If such a CFT operator exists, it will show that the region immediately beyond the inner horizon, which lies in the `future copy' of the spacetime, is also dual to the CFTs on the two boundaries of the `past copy'. But \cite{Papadodimas:2019msp} have identified a necessary condition for the smoothness of the inner horizon in terms of bulk correlation functions between fields on two sides of the inner horizon. In the CFT, this would translate to a condition on the correlation function between the boundary representations of fields inside and outside the inner horizon. The former was obtained in this paper. Any candidate for the latter would thus have to satisfy certain (quite restrictive) conditions on their correlation functions with the boundary representations given here. 

Another possible application arises in the context of the paper \cite{Balasubramanian:2019qwk}. In this paper, the stability of the inner horizon under perturbation by a shock-wave was studied by inserting a heavy local operator and studying particular CFT correlators which are indicative of the stability of the horizon. While this study was indicative, one can obtain a more complete picture from the CFT by looking at how the boundary representations of bulk fields near the inner horizon evolve under a shock wave perturbation. Indeed, general perturbations of the inner horizon can be studied from the CFT from the boundary representations obtained here.

\acknowledgements{We thank Partha Paul for his initial collaboration and significant contribution to the paper. Without his contributions, this paper would not have been possible. We also thank Parijat Dey for the helpful discussions. }

\end{document}